# Nonlocally coupled moisture model for convective self-aggregation

[§] Tomoro Yanase,[a,b,c] Shin-ichiro Shima,[c,b] Seiya Nishizawa,[b] and Hirofumi Tomita[b,a]

[a] *RIKEN Cluster for Pioneering Research, Kobe, Japan*

[b] *RIKEN Center for Computational Science, Kobe, Japan*

[c] *Graduate School of Information Science, University of Hyogo, Kobe, Japan*

§ Current affiliation: Graduate School of Information Science, University of Hyogo, Kobe, Japan

*Corresponding author*: Tomoro Yanase, yanase@gsis.u-hyogo.ac.jp

## ABSTRACT

Clouds play a central role in climate physics by interacting with precipitation, radiation, and circulation. Despite being a fundamental issue in convective organization, the self-aggregation of clouds lacks a theoretical explanation due to its complexity. In this study, we introduce an idealized mathematical model where the system's state is represented solely by the vertically integrated water vapor content of atmospheric columns under the weak temperature gradient approximation. By analyzing the nonlinear dynamics of this simplified system, we mathematically elucidate the mechanisms that determine the onset of self-aggregation and the spatial scale of the self-aggregated state. Nonlocal coupling between atmospheric columns induces bistability, leading to dry and moist equilibria. This reflects the circulation effects driven by horizontal differential heating due to convection and radiation. The bistable self-aggregated state realizes when destabilization by nonlocal coupling, triggered by finite-amplitude disturbances in the uniform state, overcomes stabilization by diffusion. For globally coupled systems where all columns are equally coupled, perturbations with the maximum wavelength exhibit the highest growth rate. This results in a solution with an infinitely long wavelength, understood as the dynamical system's heteroclinic trajectories describing the steady state's spatial evolution. Conversely, for nonlocally coupled systems with finite filter lengths, perturbations with wavelengths close to the characteristic length of the coupling are preferred. The results reveal that the balance between nonlocal coupling and diffusion is essential for understanding convective self-aggregation. Moreover, this study suggests a physical similarity between convective self-aggregation and moisture mode.

## SIGNIFICANCE STATEMENT

Cloud self-organization is a longstanding fundamental problem in climate physics, and its representation in climate models may contribute to uncertainties in future climate projections. Clouds are complex phenomena, intricately connected to latent heat release during water phase changes, buoyancy-driven fluid motion, and interactions with radiation. As a result, their detailed modeling is ongoing. Efforts have also been undertaken to understand the macroscopic behavior of clouds through simple mathematical descriptions. In this study, we semi-analytically reveal the mechanisms underlying the spontaneous clustering of clouds and the characteristic distances between clusters using an idealized mathematical model that describes the spatiotemporal variation of water vapor content in tropical atmospheric columns.

# 1. Introduction

Clouds play a central role in determining Earth's climate (Siebesma et al. 2020). Among the various cloud phenomena, the roles and mechanisms of organized clouds are not well understood (Neelin et al. 2008; Sherwood et al. 2010; Bony et al. 2015; Mauritsen and Stevens 2015; Reed and Medeiros 2016; Sherwood et al. 2020). Recently, convective self-aggregation (CSA), the spontaneous clustering of convective clouds, has attracted attention, as it directly affects climate systems (Held et al. 1993; Tompkins and Craig 1998; Bretherton et al. 2005; Wing et al. 2017). However, a comprehensive theoretical understanding of its mechanism and characteristics has not yet been reached, and its relationship with real atmospheric phenomena is not yet understood (Muller et al. 2022).

CSA theory should encompass the following features that have been revealed primarily through numerical studies using complex models such as cloud-resolving models (CRMs) and general circulation models in the framework of radiative–convective equilibrium (RCE):

1. The system possesses two states, including a scattered state in which clouds are randomly generated in a uniform field and an aggregated state in which the dry descending clear area and moist ascending cloudy area are separated.
2. A critical domain size of a few hundred kilometers exists for the onset, below which the state is scattered and above which the state is aggregated (Muller and Held 2012; Jeevanjee and Romps 2013; Yanase et al. 2020).
3. The inter-cluster distance exhibits a characteristic length of a few thousand kilometers, and a structure of that length is selected over a sufficiently large region exceeding the order of ten thousand kilometers (Wing and Cronin 2016; Patrizio and Randall 2019; Matsugishi and Satoh 2022; Yanase et al. 2022b).
4. Physical feedback between thermodynamics and dynamics such as water vapor, clouds, radiation, surface flux, and circulation play a role. However, which of these feedbacks is essential is a matter of debate, and it is also considered to depend on external environments (e.g., sea surface temperature) (SST; Emanuel et al. 2014; Wing and Emanuel 2014; Coppin and Bony 2015; Wing et al. 2020; Yanase et al. 2022a; Yao and Yang 2023).

Several studies have attempted to understand CSA using simple models. A class of mathematical models is based on ordinary differential equations for the local dynamics of the atmospheric column (Bretherton et al. 2005; Emanuel et al. 2014). Another class is based on partial differential equations for spatiotemporal dynamics of horizontally extended atmospheric fields. Two representative examples are the reaction–diffusion (RD) model (Craig and Mack 2013; Windmiller and Craig 2019; Biagioli and Tompkins 2023) that describes the time evolution of the water vapor field and the gravity wave (GW) model (Yang 2021) that incorporates a mass source and sink mimicking radiative cooling and convective heating, respectively, in a shallow water dynamic equation system.

Studies based on the water vapor RD model have focused on a coarsening process to explain the onset and initial development of CSA. These studies highlight the importance of bistable potentials in the reaction terms and the existence of diffusion. Windmiller and Craig (2019) adopted the local simple models of CSA (i.e., Bretherton et al. 2005; Emanuel et al. 2014) in constructing a reaction term to indicate the similarity of pattern formation despite the different local instability mechanism. This approach also led to an attempt to interpret the time evolution of the horizontal distribution of water vapor in the tropical convergence zone as a dynamical system with bistable potential (Beucler et al. 2020). However, unlike the results of the large-domain RCE simulation using CRMs (Matsugishi and Satoh 2022; Yanase et al. 2022b), the RD model does not exhibit the coexistence of multiple clusters, as it lacks a mechanism to determine the typical scale of cloud clusters. Furthermore, it does not include advection term and dynamic interactions through the atmospheric circulation field, missing the third and fourth properties mentioned earlier.

Conversely, the GW model emphasizes that the spatial scale of the adjustment of the atmospheric circulation field due to GWs (Bretherton and Smolarkiewicz 1989) plays a role in the scale selection of CSA. In that model, radiative cooling increases the geopotential as a horizontally uniform mass source, while convective heating is triggered when the geopotential exceeds a certain threshold and decreases the geopotential as a mass sink in the finite temporal and horizontal range. Although the coupling of thermodynamics and dynamics is expressed via the geopotential, there is no feedback mechanism that allows the field to alter the radiative effect. While many studies have reported a role of radiative interactions in the CSA onset (Tompkins and Craig 1998; Bretherton et al. 2005; Muller and Held 2012; Yang 2018), the GW model does not represent that dependence. Hence, it cannot handle thermodynamic interactions despite the existence of one-way effect from constant radiation to geopotential, lacking the fourth property.

We sought to determine if we can build a unified model that incorporates the strengths of both existing models and compensates for their deficiencies. A clue to answer this question is obtained by investigating the interaction between large-scale circulation fields and tropical convection based on *weak temperature gradient* (WTG) approximation (Sobel and Bretherton 2000; Raymond and Zeng 2005). The WTG framework is used to model a physical situation in which a quasi-local moist convective region is coupled to a large-scale environmental region via GW adjustment that homogenizes the temperature field horizontally. Several previous studies have reported the existence of multiple equilibrium states in a single column model (Sobel et al. 2007) and a small-domain CRM (Sessions et al. 2010) based on the WTG framework. This suggests that the atmospheric column can be regarded as a dynamical system with a bistable potential under the WTG approximation as suggested by Windmiller and Craig (2019). The WTG framework possesses the potential to incorporate the physical essences of both bistable RD model and GW model. However, it is unclear what horizontal structure is created by the horizontally interacting atmospheric columns in the WTG framework.

From this point of view, we focus on the water vapor reaction–diffusion–advection model proposed by Ahmed and Neelin (2019, AN19). Although the primary purpose of AN19 was to explain the size distribution of tropical mesoscale precipitation systems, they also reported the occurrence of CSA under certain conditions. However, they did not explain the mechanism of CSA onset or the horizontal scales associated with it. The strength of the AN19 model lies in its inclusion of a feedback mechanism between thermodynamics and dynamics. This is achieved through the coupling of heating and circulation as a diagnostic equation of dry static energy under the WTG approximation while maintaining the prognostic equation of water vapor at its core. However, the heating–circulation coupling in the AN19 model is formulated in an extreme form of "global coupling", where one atmospheric column is equally coupled to all other atmospheric columns. As the WTG approximation is based on the horizontal homogenization effect of the temperature field due to GWs, this implies that the effect of GWs extends instantaneously over an infinite distance. The scaling relationship proposed by the GW model of Yang (2021) predicts that the spatial scale of CSA diverges to infinity. Conversely, we speculate that altering the heating–circulation coupling in the AN19 model to a more natural and general form of "nonlocal coupling" (Shima and Kuramoto 2004), in which the coupling strength between one point and another depends on the distance would allow for scale selection of a finite CSA.

In this study, we investigate the occurrence of CSA by examining the extended nonlocally coupled AN19 equation in greater detail, both theoretically and numerically. After introducing the mathematical model in Section 2, we address whether the wavelength of CSA can be infinitely extended in a globally coupled system and whether transitioning to a nonlocally (but not globally) coupled system results in the selection of a specific wavelength. This is explored through numerical experiments with the extended AN19 equation (Section 3) and stability analysis of the uniform state (Section 4). Additionally, the infinite wavelength solution in a globally coupled system is mathematically examined using a dynamical systems approach (Section 5).

Through these analyses, we propose a unified understanding of CSA occurrence and scale selection by integrating essential aspects of both the water vapor RD model and the GW model frameworks. Furthermore, extending the AN19 equation to CSA theory yields implications for understanding CSA in relation to the Madden–Julian Oscillation (MJO) and water vapor variability in a natural tropical atmosphere. The AN19 model shares a common physical basis with the moisture mode theory (Sobel and Maloney 2012, 2013), one of the leading theories of the MJO. Attempts have been made to compare the nature of water vapor variability between actual and idealized atmospheres based on a similar equation (Beucler et al. 2019). In Section 6, we discuss the relationship between our theory and these studies. Conclusions are provided in Section 7.

## 2. Nonlocally coupled moisture model

This study primarily utilizes the time evolution equation for the horizontally extended field of vertically integrated water vapor content, $q_v$. This equation is based on the system of equations developed by AN19 with some simplifications and extensions. Details of the derivation and assumptions are provided in the Appendix. The set of equations is:

$$\frac{\partial q_v}{\partial t} = E - \mathcal{P} + M_q(\nabla \cdot \boldsymbol{v})q_v + M_q(\boldsymbol{v} \cdot \nabla)q_v + D\nabla^2 q_v \,, \tag{1}$$

$$\nabla \cdot \boldsymbol{v} = \frac{L_v\left(\mathcal{P} - \tilde{\mathcal{P}}\right) + \left(\mathcal{F}_{\mathrm{rad}} - \tilde{\mathcal{F}}_{\mathrm{rad}}\right) - \xi}{M_s} \,, \tag{2}$$

where $E$ is the evaporation rate constant; $D$ is the eddy diffusion coefficient constant; $M_q$ is the gross moisture stratification constant per unit $q_v$; $M_s$ is the gross dry stability constant; $L_v$ is the latent heat constant; and $\boldsymbol{v}$ is the irrotational horizontal wind vector in the upper troposphere. The vertical structures of the large-scale vertical velocity and water vapor content are represented by single modes of the first baroclinic-type and exponential-type functions, respectively (see Appendices B and D for details). The term $\mathcal{P}$ represents precipitation, $\mathcal{F}_{\mathrm{rad}}$ denotes the column radiative heating term, and $\xi$ is a stochastic term, as formulated later.

The expression $\tilde{A}$ denotes the horizontally filtered value of a variable $A$, as defined by $\tilde{A}(x) = \int_{-\infty}^{\infty} G(r)A(x+r)\,\mathrm{d}r$, where $G(r)$ is the filter function. This represents a natural extension of the operation originally treated as a horizontal average over the domain in the AN19 equation to a more general horizontal filter. This filter operation physically defines the horizontal scale at which the WTG approximation is valid for the thermodynamic equations, and the horizontal scale can be determined by the adjustment process of the dynamics. In other words, here we assume that the filtered divergence vanishes, which effectively means we are prescribing the length scale of the self-organized circulation.

For instance, within the context of moisture mode theory for the MJO, Sobel and Maloney (2012, 2013) used a filter function with east-west asymmetry to reflect the differences in the propagation characteristics of Kelvin and Rossby waves on the equatorial beta plane when diagnosing the circulation field from the heating field. In this study, assuming a non-rotating atmosphere, an isotropic (symmetric) filter function is employed. For simplicity, we use a box filter:

$$G(r) = \mathcal{H}(l_{\mathrm{B}}/2 - |r|)/l_{\mathrm{B}},$$

where $l_{\mathrm{B}}$ denotes the filter length, and $\mathcal{H}(A)$ is the Heaviside step function defined as $\mathcal{H}(A) = 0$ for $A \leq 0$ and $\mathcal{H}(A) = 1$ for $A > 0$. The filter length $l_{\mathrm{B}}$ represents a horizontal scale indicative of the effective propagation distance of GWs. Although this treatment is simple, it incorporates an important aspect of the dynamical effect of GW adjustment into the water vapor equation. Notably, this formulation generally defines the system as a nonlocally coupled system. In special cases, it is referred to as a globally coupled system when $l_B$ is infinite or equal to the domain size.

The precipitation term $\mathcal{P}$ and the column radiative heating term $\mathcal{F}_{\mathrm{rad}}$ were parameterized as follows:

$$\mathcal{P} = \alpha(q_v - q_c)\mathcal{H}(q_v - q_c)\,, \tag{3}$$

$$\mathcal{F}_{\mathrm{rad}} = \epsilon_{\mathrm{r}} q_v + \mathrm{const.}\,, \tag{4}$$

where $\alpha$ is the coefficient of relaxation by precipitation, and $q_c$ is the critical water vapor content (the prescribed value is 40 kg m$^{-2}$ in this study). The parameter $\epsilon_r$ represents the coefficient of the water vapor radiative effect. Here, as $\mathcal{F}_{\mathrm{rad}}$ is meaningful only as a deviation from the filtered value in Eq. (2) in our model, only the dependence on $q_v$ is important for the radiation parameterization, and the specific value of the offset constant is not important. The stochastic term $\xi$ is modeled as the Ornstein–Uhlenbeck process (Gardiner 2004):

$$\mathrm{d}\xi = -\tau_{\mathrm{noise}}^{-1}\xi \mathrm{d}t + \tau_{\mathrm{noise}}^{-1}\sigma_{\mathrm{noise}}\mathrm{d}W(t)\,, \tag{5}$$

where $\tau_{\mathrm{noise}}$ is the damping time scale, $\sigma_{\mathrm{noise}}$ is the intensity of unresolved variability, and $\mathrm{d}W(\mathrm{t})$ denotes the Wiener process. Physically, this term expresses the effect of subgrid variability, such as high-frequency GWs, that cannot be structurally represented by the water vapor equation and WTG thermodynamic equation alone.

Figure 1 presents a schematic of the model. Formally, the system is composed of a reaction term $\mathcal{R}$, advection term $\mathcal{A}$, diffusion term $\mathcal{D}$, and stochastic term $\mathcal{X}$:

$$\frac{\partial q_v}{\partial t} = \mathcal{R} + \mathcal{A} + \mathcal{D} + \mathcal{X}\,, \tag{6}$$

$$\mathcal{R} = E - \mathcal{P} + M_q \frac{L_v\left(\mathcal{P} - \tilde{\mathcal{P}}\right) + \epsilon_{\mathrm{r}}(q_v - \widetilde{q_v})}{M_s} q_v\,, \tag{7}$$

$$\mathcal{A} = M_q(\boldsymbol{v} \cdot \nabla)q_v\,, \tag{8}$$

$$\mathcal{D} = D\nabla^2 q_v\,, \tag{9}$$

$$\mathcal{X} = -\frac{M_q}{M_s}\xi q_v\,. \tag{10}$$

This classification follows previous studies using the RD equation as a mathematical model of CSA (Craig and Mack 2013; Windmiller and Craig 2019; Biagioli and Tompkins 2023). Our model differs from those studies by the inclusion of an advection term in our model. Additionally, it is notable that $\mathcal{R}$ is a function of both the local value $q_v$ and the filtered value $\widetilde{q_v}$ (and $\tilde{\mathcal{P}}$); therefore, we refer to this model as a nonlocally coupled moisture model.

Before discussing the entire system, let us first examine the basic properties without horizontal interaction (i.e., $\mathcal{A}$ and $\mathcal{D}$) and stochasticity ($\mathcal{X}$) to focus on the equilibrium states achievable by local atmospheric column dynamics driven solely by $\mathcal{R}$. The reduced dynamical system equation can be written as:

$$\frac{\mathrm{d}q_v}{\mathrm{d}t} = -\frac{\mathrm{d}V(q_v)}{\mathrm{d}q_v},$$

where $V(q_v)$ is the potential function defined as follows:

$$V(q_v) = -\int_0^{q_v} \mathcal{R}(q_v')\,\mathrm{d}q_v' + \text{const.}$$

As

$$\frac{\mathrm{d}V}{\mathrm{d}t} = \frac{\mathrm{d}V}{\mathrm{d}q_v}\frac{\mathrm{d}q_v}{\mathrm{d}t} = -\left(\frac{\mathrm{d}V}{\mathrm{d}q_v}\right)^2 \leq 0,$$

$V$ decreases monotonically with time or stays at a fixed point where $\mathrm{d}V/\mathrm{d}q_v = 0$ (Strogatz 2019). Therefore, for the dynamics of the atmospheric column in our model, it is essentially important what form $V$ takes.

Figure 2a presents potential functions (defined above and explained in detail in the Appendix), and Figure 2b provides a bifurcation diagram representing the local equilibrium point of $q_v$ when $\widetilde{q_v}$ is considered a controlled parameter. Here, we assumed $\tilde{P} = E$. Although this assumption is ad-hoc and arbitrary, it does not affect the qualitative explanation for demonstrative purposes below. As presented in Figure 2a, the shape of the potential changes with the value of $\widetilde{q_v}$. When $\widetilde{q_v}$ is 5 kg m$^{-2}$ or 45 kg m$^{-2}$, there is a single potential well (i.e., a stable equilibrium solution) in the displayed range. In contrast, when $\widetilde{q_v}$ is 25 kg m$^{-2}$, there are two potential wells, one on the moist side and the other on the dry side. Figure 2b summarizes the existence of the well (stable fixed point) and ridge (unstable fixed point) of potential as a function of $\widetilde{q_v}$. As indicated by the changes in the potential functions and the bifurcation of the local equilibrium point of $q_v$ with respect to variations in $\widetilde{q_v}$ within the displayed range, this system can exhibit bistability. Specifically, two stable solutions can coexist for a specific range of $\widetilde{q_v}$. We note that the intersection of the solid line and the line $q_v = \widetilde{q_v}$ in Figure 2 indicates the presence of a linearly stable horizontally uniform solution in the original horizontally extended system. Notably, the uniformity of precipitation is also ensured as noted above. Finding the solution of $\mathcal{R} = 0$ under the conditions $q_v = \widetilde{q_v}$ and $\mathcal{P} = \tilde{\mathcal{P}}$, the uniform solution $q_{v0}$ is given by:

$$q_{v0} = q_\mathrm{c} + \frac{E}{\alpha}\,, \tag{11}$$

where precipitation and evaporation are balanced in each column (i.e., $\mathcal{P} = E$) The growth rate $\sigma$ for linear perturbations in the form of $\propto \exp(\sigma t + ikx)$ about the uniform state is:

$$\sigma = -\alpha + \frac{M_q}{M_s} q_{v0}(L\alpha + \epsilon_\mathrm{r})\left(1 - \hat{G}(k)\right) - Dk^2, \tag{12}$$

where $k$ is the horizontal wavenumber, and $\hat{G}(k)$ is the transfer function defined as $\hat{G}(k) = \int_{-\infty}^{\infty} G(r)e^{ikr}\,\mathrm{d}r$. When using a box filter, $\hat{G}(k) = \sin(kl_{\mathrm{B}}/2)/(kl_{\mathrm{B}}/2)$. As $\sigma < 0$ for the parameters used in this study, the uniform state is linearly stable.

The question is whether this uniform solution, when subjected to a finite-amplitude disturbance, remains in a uniform state (i.e., a scattered state) or transitions to an aggregated state in which two stable solutions (dry and moist) are achieved. If a transition occurs, what wavelength is selected? In the next section, we address this issue using numerical experiments and theoretical considerations of the extended AN19 model.

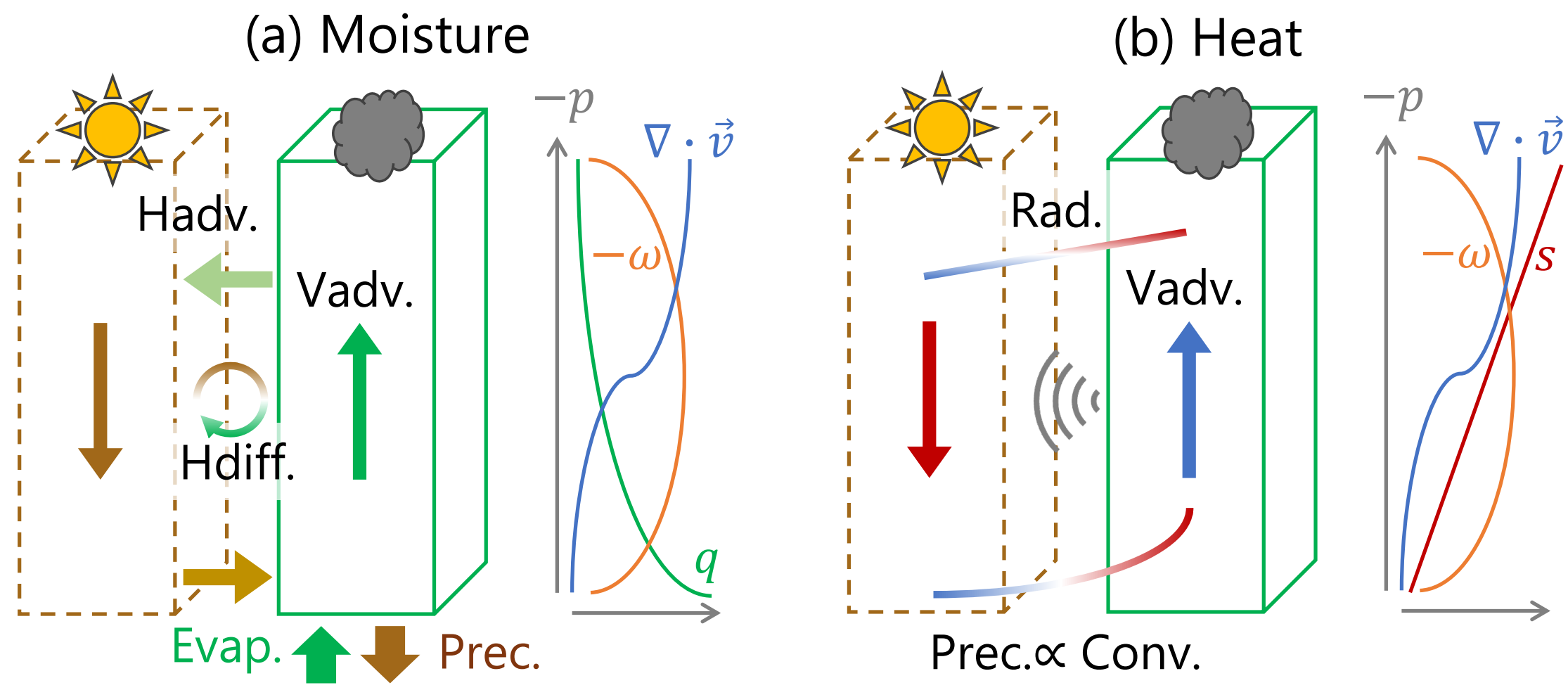

Fig. 1. Schematic of the moisture model. (a) The budget equation for moisture leads to the prognostic equation for vertically integrated water vapor content that includes surface evaporation (Evp.), surface precipitation (Prec.), vertical advection (Vadv.), horizontal advection (Hadv.), and horizontal eddy diffusion (Hdiff.). (b) The budget equation for heat results in the diagnostic equation for vertically integrated dry static energy, incorporating convective heating (Conv.; proportional to surface precipitation), radiative heating (Rad.), and vertical advection (Vadv.). Vertical structures of vertical velocity, water vapor content, and dry static energy are represented by half sine wave functions, exponential functions, and linear functions, respectively.

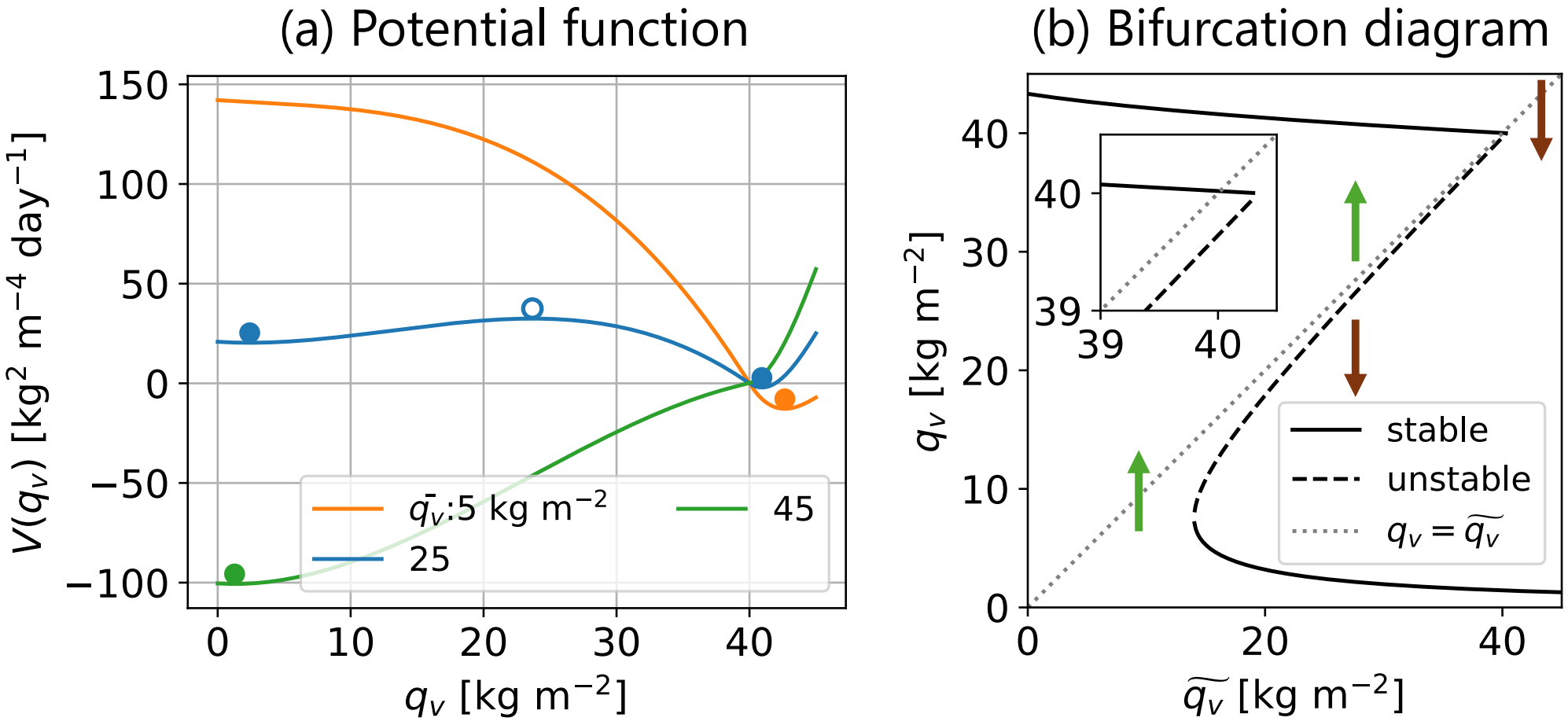

Fig. 2. Dynamical features of the reaction term. (a) Potential function $V(q_v)$ for different $\widetilde{q_v}$. Filled and open circles represent stable and unstable fixed points, respectively. (b) Bifurcation diagram of $q_v$ against $\widetilde{q_v}$. Solid and dashed lines represent stable and unstable fixed points, respectively. The dotted line represents $q_v = \widetilde{q_v}$. The water balance $\tilde{P} = E$ is assumed. Green and brown arrows indicate the direction of change (i.e., moistening and drying tendency, respectively).

## 3. Numerical experiments

First, we performed a series of numerical experiments to investigate the time evolution of the extended AN19 model. The horizontal discretization was based on a Cartesian coordinate system with periodic lateral boundary conditions. The advection term and divergence term (which arise from the vertical integration of horizontal advection and vertical advection, respectively) were computed simultaneously as the horizontal flux divergence, combined into the flux form. The velocity vector field was determined as the gradient of the velocity potential. The scalar field of the velocity potential was calculated by expressing the velocity divergence term in the thermodynamic equation in terms of the Laplacian of the velocity potential and numerically solving the resulting Poisson equation using the successive over-relaxation method. The horizontal flux divergence was computed using the second-order centered difference method, with the horizontal flux at a point shifted by half a grid from the position of the scalar value. The time evolution of the stochastic term was computed using the Euler–Maruyama method, with the Wiener process represented by normal random numbers generated using the Box–Muller method. The horizontal grid size was set to 20 km, and the time interval was 300 s. The initial column water vapor field was set to a uniform value of 45 kg m$^{-2}$ plus uniform random numbers with a width of 1 kg m$^{-2}$.

We conducted global coupling runs with different domain sizes of 640, 2,560, and 10,240 km and nonlocal coupling runs with different filter lengths of 1,500, 3,020, and 6,020 km on the 10,240 km domain using a 1D setup. The time integration period was 500 days for the 10,240 km domain runs and 1,000 days for the smaller domain runs. Figure 3 indicates the horizontal–time distribution of $q_v$ obtained from the numerical experiments. As depicted in Fig. 3a, for the global coupling run with a 640 km domain, the $q_v$ field remained nearly uniform, indicating a scattered state where precipitation occurred everywhere. As the domain size increased, the $q_v$ field transitioned into an aggregated state with distinct dry and moist regions (Figs. 3b and 3c). For the global coupling system, only a single isolated moist cluster existed in the periodic domain. Hence, it could be stated that the horizontal distance between the moist cluster is restricted and determined by the domain size. If the domain size were increased to infinity, it remains uncertain whether the inter-cluster distance would also increase to infinity. Due to the challenges in determining system behavior at infinite domain sizes through numerical experiments, we discuss the system's structure in the absence of domain constraints mathematically from the perspective of dynamical systems bifurcation in Section 5. Although the horizontal propagation of cluster as observed in Figure 3c is an interesting phenomenon, the scope of this study is limited to horizontal scale selection, and we leave the propagation problem as a future issue.

In contrast, multiple moist clusters coexisted in the nonlocal coupling runs, as presented in Figs. 3d–f, with the horizontal distance between moist clusters increasing as the filter length increased. The scale selection mechanism is discussed in the following section.

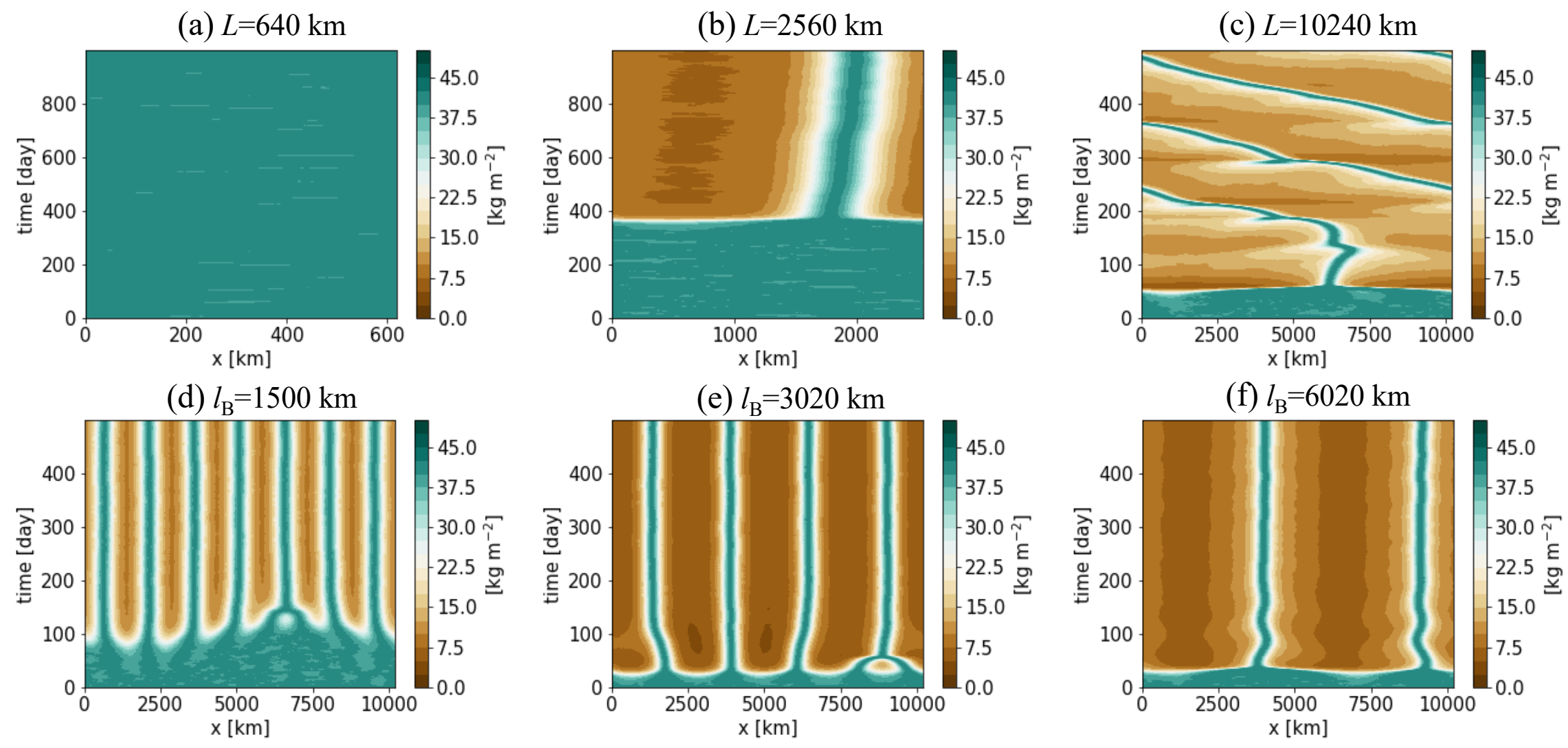

Fig. 3 Horizontal–time distribution of $q_v$. (a)–(c) Global coupling runs on 640, 2,560, and 10,240 km domains. (d)–(f) Nonlocal coupling runs with 1,500, 3,020, and 6,020 km filter lengths on the 10,240 km domain.

## 4. Stability of uniform state: Onset and scale selection

We now theoretically examine why specific wavelengths are selected for nonlocally coupled systems in terms of the stability of the uniform solution. The uniform solution is linearly stable (Eq. (12)), as indicated by the potential function corresponding to the linearized reaction term (orange line in Fig. 4a). Therefore, it is essential to incorporate the effects of finite-amplitude disturbances to address the onset and scale selection of CSA. In this context, we focus on the nonlinearity of the precipitation term that arises due to the Heaviside function when considering significant drying anomalies from a uniform solution, as presented in Fig. 4b.

We propose that when $q_v$, given a (drying) perturbation $q_v'$ from $q_{v0}$, falls below $q_c$ (i.e., $q_v' = q_v - q_{v0} < q_\mathrm{c} - q_{v0}$), the precipitation effect $\alpha$ changes to the effective precipitation effect:

$$\alpha_\mathrm{eff} = \alpha \frac{(q_{v0} - q_\mathrm{c})}{\Delta q_v} , \tag{13}$$

where $\Delta q_v = |q_v'|$ is the absolute value of the dry perturbation (green circle in Fig. 4b). The potential function corresponding to the reaction term using the effective precipitation effect (green line in Fig. 4a) can capture the existence of a ridge (local maximum) located on the dry side of the valley (local minimum) corresponding to the uniform state, resembling the original nonlinear version (blue line in Fig. 4a).

By using $\alpha_\mathrm{eff}$, a new effective growth rate can be defined as:

$$\sigma_\mathrm{eff} = -\alpha_\mathrm{eff} + \frac{M_q}{M_s} q_{v0} (L\alpha_\mathrm{eff} + \epsilon_\mathrm{r}) \left(1 - \hat{G}(k)\right) - Dk^2 . \tag{14}$$

Fig. 5a indicates the effective growth rate as a function of the wavenumber for both globally coupled and nonlocally coupled systems with two different $l_\mathrm{B}$ values, assuming a 10,240 km domain. Here, $\Delta q_v = 1\ \mathrm{kg\ m^{-2}}$ is used as a typical scale (the conclusion does not change even if the value is increased). Contributions from precipitation (the first term on the right-hand side [rhs) of Eq. (14)), divergence term (the second term), and diffusion (the third term) are presented separately.

In a globally coupled system (gray line in Fig. 5a), the effective growth rate reaches its maximum positive value at the longest wavelength of 10,240 km (the smallest wavenumber; gray circle in Fig. 5a) and decreases as the wavelength shortens (the wavenumber increases). As the divergence and precipitation terms do not vary with the wavenumber, the change in the effective growth rate is due solely to diffusion. If the domain size increases, the wavelength with the maximum effective growth rate also increases.

In the nonlocally coupled system with a 2,560 km filter (blue line in Fig. 5a), the effective growth rate achieves its maximum positive value at an intermediate wavelength of 2,048 km (blue circle in Fig. 5a). The difference from the globally coupled system is attributed to the longwave cutoff of the divergence term caused by nonlocal coupling (blue dashed line in Fig. 5a). With the 640 km filter (orange line in Fig. 5a), the effective growth rate is negative at all wavelengths, as the diffusion term dominates.

To provide an overall view of the unstable mode, Fig. 5b displays the effective growth rate as a function of filter length and wavelength. It also indicates the wavelength with the maximum growth rate for a given filter length, along with results obtained from numerical experiments using the extended AN19 model. The following observations can be made. (1) The region with a positive effective growth rate is generally confined to wavelengths longer than approximately 1,000 km, and (2) The wavelength with the maximum effective growth rate is located near the filter length (close to the line $y = x$). The first observation is related to the CSA onset problem and is primarily due to diffusion that is proportional to the square of the wavenumber. The second observation pertains to the scale selection of the inter-cluster distance and is primarily caused by the longwave cutoff induced by nonlocal coupling. These two features can be further examined by considering an approximated situation where $\alpha_{\mathrm{eff}} = 0$.

Regarding the first feature, the critical wavenumber $k_{\mathrm{c1}}$ that represents the boundary between unstable and stable modes (i.e., $\sigma_{\mathrm{eff}}(k_{\mathrm{c1}};\ \alpha_{\mathrm{eff}} = 0) = 0$), satisfies the equation:

$$\left(\frac{M_q}{M_s}\right) q_{v0}\epsilon_{\mathrm{r}} \left(1 - \hat{G}(k_{\mathrm{c1}})\right) - D{k_{\mathrm{c1}}}^2 = 0\,,$$

from Eq. (14). Additionally, if the scale of interest is sufficiently shorter than the filter length or if the system is globally coupled, then $\hat{G} = 0$. In such cases, $k_{\mathrm{c1}}$ (considering only the positive root) can be analytically expressed by the system parameters:

$$k_{c1} = \sqrt{\frac{\epsilon_r M_q}{D M_s}\left(q_c + \frac{E}{\alpha}\right)}\,. \tag{15}$$

The critical wavelength corresponding to $k_{\mathrm{c1}}$ is presented in Fig. 5b. This expression indicates that a smaller diffusion coefficient or a larger water vapor radiative effect will result in a larger critical wavenumber, facilitating the onset of CSA for a given wavenumber. It could be predicted that with all else being equal, a model that exhibits stronger water vapor radiation effects will be able to generate CSA in a smaller horizontal computational domain.

Conversely, concerning the second feature, the critical wavenumber $k_{\mathrm{c2}}$ that exhibits the maximum effective growth rate (i.e., $\mathrm{d}\sigma_{\mathrm{eff}} / \mathrm{d}k(k_{\mathrm{c2}};\ \alpha_{\mathrm{eff}} = 0) = 0$), satisfies the equation:

$$-\left(\frac{M_q}{M_s}\right) q_{v0}\epsilon_{\mathrm{r}} \left(\frac{\mathrm{d}\hat{G}(k_{\mathrm{c2}})}{\mathrm{d}k}\right) - 2Dk_{\mathrm{c2}} = 0.$$

This expression indicates that $k_{\mathrm{c2}}$ is determined by the balance between both terms. From Fig. 5a, it can be observed that the filter scale is the primary determinant in the present conditions. Nevertheless, diffusion also plays a role in selecting a sharper maximum growth wavenumber.

At the end of this section, we note that the position of the maximum effective growth rate aligns well with the results obtained from the numerical experiments using the extended AN19 model with full nonlinearity. Physically, the filter scale reflects the spatial scale at which GW adjustment is effective and should be provided as an external parameter in our model. Assuming the horizontal propagation speed of the first baroclinic mode to be 50 m $\mathrm{s}^{-1}$ (Ruppert and Hohenegger 2018) and the damping time scale to be one day (Yang 2021, 2018), the horizontal length scale, calculated as the product, is approximately 4,000 km. At this filter scale, the wavelength corresponding to the maximum effective growth rate was approximately 3,000 km. This scale is consistent with the results of Yanase et al. (2022b), who used a complex CRM and observed that the characteristic distance of CSA was 3,000–4,000 km. Additionally, we confirmed that the wavelength with the maximum effective growth rate can be obtained even when using a Gaussian filter. This result indicates that a spatial structure with such a wavelength is selected in the numerical experiments, although a box filter was used in this study.

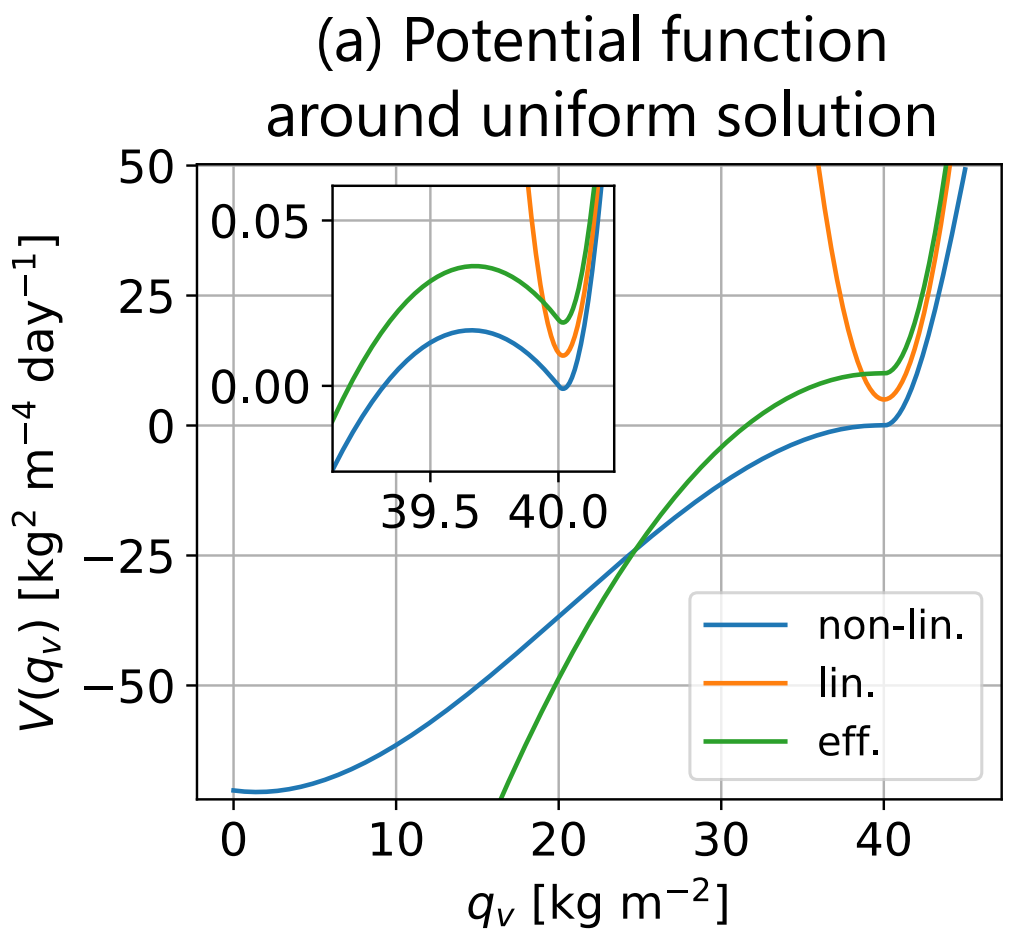

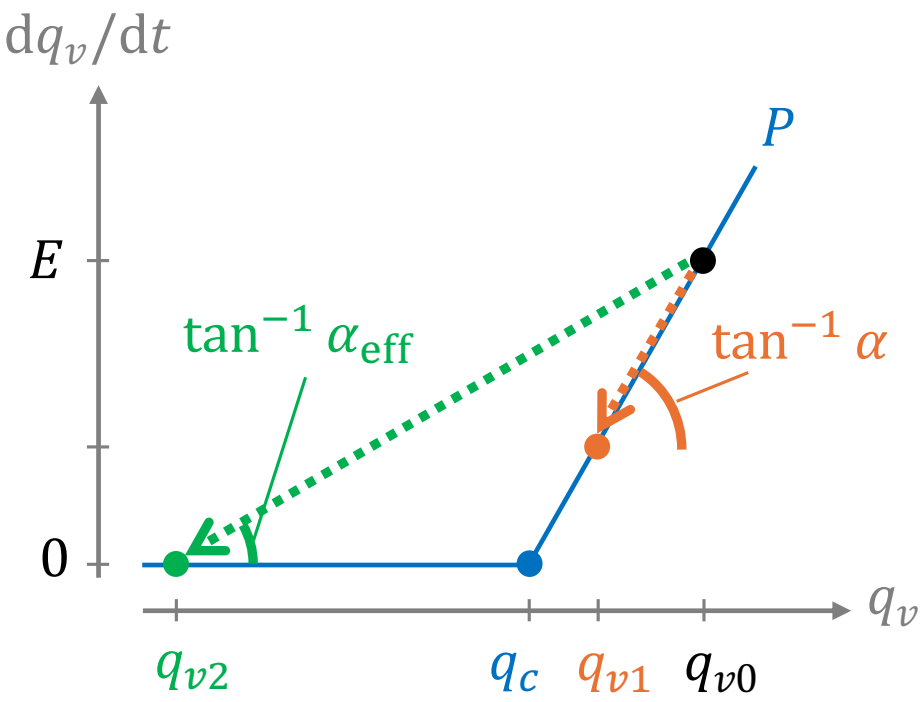

Fig. 4. Potential function and precipitation effect around the uniform solution. (a) Potential function $V(q_v)$ around the uniform solution. Blue, orange, and green lines represent the potential function based on the original nonlinear, linearized, and effective precipitation terms, respectively. Lines are vertically shifted for visibility. (The value of the potential is not necessarily zero at $q_v = q_c$, but a small value is added. There is no change in the physical meaning.) (b) Schematic of effective precipitation. In the phase space $(q_v, \mathrm{d}q_v/\mathrm{d}t)$, the precipitation term for the uniform solution is indicated by the black circle at $(q_{v0}, E)$. Precipitation terms perturbed negatively small and large from the uniform solution are indicated by the orange and green circles at $(q_{v1}, E - \alpha(q_{v0} - q_{v1}))$ and $(q_{v2}, 0)$, respectively. The precipitation effect (inverse time scale) is $\alpha$ when $q_v$ after perturbation is larger than $q_c$, whereas it is $\alpha_{\mathrm{eff}}$ (see Eq. (13)) depending on the magnitude of the perturbation when $q_v$ after perturbation is smaller than $q_c$.

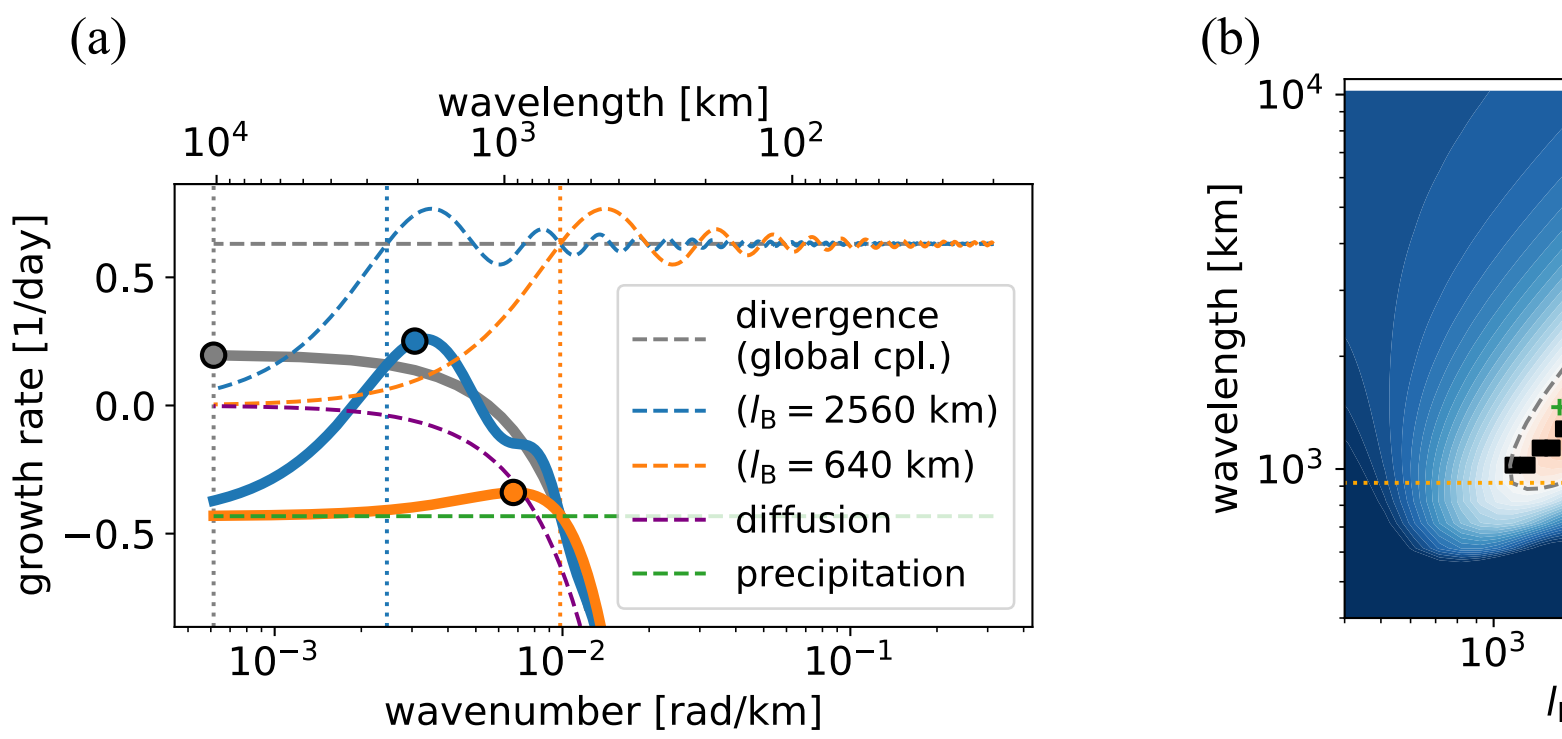

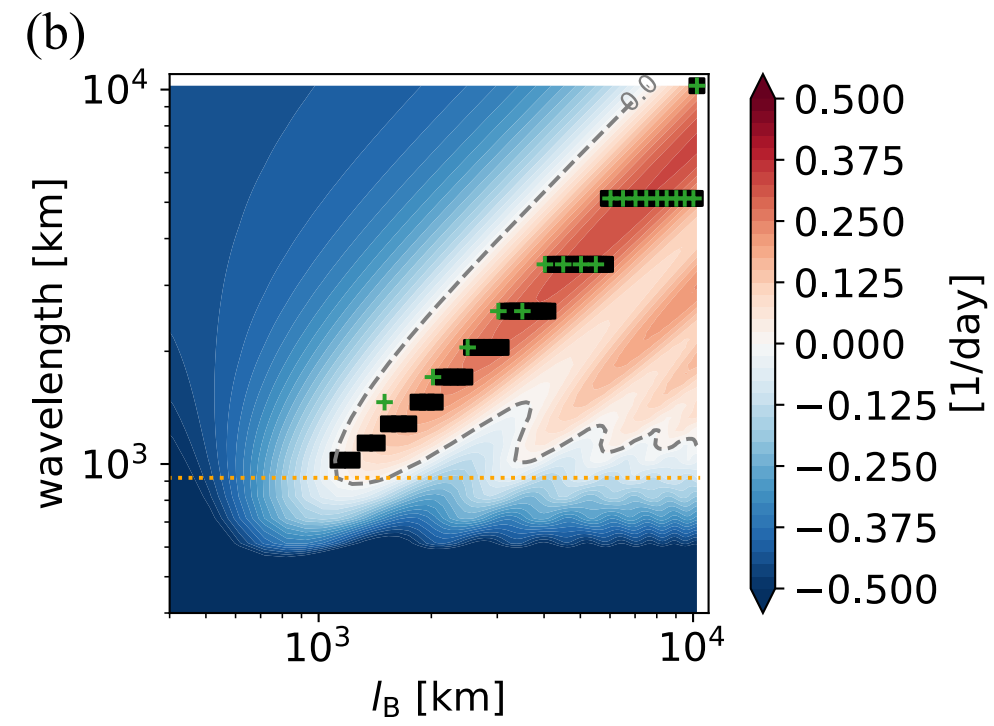

Fig. 5. Effective growth rate as a function of wavenumber and $l_\mathrm{B}$. The perturbation $\Delta q_v$ is assumed to be 1 kg m$^{-2}$. (a) Effective growth rate and contribution from each term as a function of the horizontal wavenumber. Gray, blue, and orange solid lines represent the effective growth rate for the globally coupled system and the nonlocally coupled system with $l_\mathrm{B} = 2{,}560$ and 640 km, respectively. Divergence terms are represented by dashed lines of the same color as that of the effective growth rate. Green and purple dashed lines represent the precipitation and diffusion terms, respectively. Circles on the solid lines indicate the maximum effective growth rate among the discrete modes for the periodic domain with a size of 10,240 km. Vertical dotted lines represent $l_\mathrm{B}$. Wavelength is presented on the horizontal axis at the top. (b) Effective growth rate in $l_\mathrm{B}$–wavelength space. Black squares represent the points with the maximum effective growth rate at a given $l_\mathrm{B}$. Green crosses represent results obtained from numerical experiments of the nonlocally coupled system. The orange horizontal dotted line represents the critical wavelength corresponding to Eq. (15).

## 5. Steady state solution of the globally coupled system

Finally, we investigate why the distance between clusters approaches infinity in the globally coupled system to highlight the necessity of nonlocal coupling for maintaining a finite inter-cluster distance. To understand the characteristics of the horizontal scale in a globally coupled system, we need to analyze the horizontal distribution of steady states while preserving the nonlinearity of the extended AN19 equation and eliminating the effect of finite domain size. We do this by examining a dynamical system that describes the horizontal variation of state variables and their periodic solutions.

The steady state of the 1D globally coupled extended AN19 model can be expressed as:

$$\begin{aligned} \frac{\mathrm{d}q_v}{\mathrm{d}x} &= w \\ \frac{\mathrm{d}w}{\mathrm{d}x} &= \frac{-R - M_q u w}{D} \\ \frac{\mathrm{d}u}{\mathrm{d}x} &= \frac{L_v(P - \overline{P}) + \epsilon_r(q_v - \overline{q_v})}{M_s} \end{aligned}, \tag{16}$$

where $w$ is the horizontal gradient of $q_v$, and $u$ is the horizontal wind velocity. $\bar{A}$ denotes the horizontal average value of the variable $A$. Here, we assume $\overline{P} = E$.

Before discussing the periodic trajectory, let us examine the fixed points of this dynamical system. The conditions for a fixed point are:

$$\frac{\mathrm{d}q_v}{\mathrm{d}x} = \frac{\mathrm{d}w}{\mathrm{d}x} = \frac{\mathrm{d}u}{\mathrm{d}x} = 0.$$

According to the first and second equations of the steady-state system, this implies $w = 0$ and $\mathcal{R} = 0$. When $q_v \leq q_c$, $\mathcal{P} = 0$ and $\mathcal{R} = E$ according to Eq. (3) and (7), then there is no fixed point, as the second condition ($\mathcal{R} = 0$) cannot be satisfied. When $q_v > q_c$, the third equation of Eq. (16) leads to:

$$q_v = \{L_v(E + \alpha q_c) + \epsilon_\mathrm{r}\overline{q_v}\}/(L_v\alpha + \epsilon_\mathrm{r}),$$

and the second equation leads to:

$$\mathcal{R} = \epsilon_\mathrm{r}\alpha(E/\alpha + q_c - \overline{q_v})/(L_v\alpha + \epsilon_\mathrm{r}).$$

Thus, when $q_v = \overline{q_v} = q_{v0}$, it satisfies $\mathcal{R} = 0$. Finally, $u$ is arbitrary. Additionally, it is noteworthy that even if $\mathrm{d}u\,/\,\mathrm{d}x \neq 0$, a trajectory can still be fixed in the 2D phase space of $q_v$ and $w$, as long as $\mathrm{d}q_v/\mathrm{d}x = \mathrm{d}w/\mathrm{d}x = 0$, indicating $w = 0$ and $\mathcal{R} = 0$. This is due to the observation that $u$ does not affect $q_v$ and $w$ in that situation.

A trajectory, representing the horizontal distribution, is calculated by integrating the dynamical system along the horizontal dimension $x$. For this integration, we employed the fourth-order Runge–Kutta method. Periodic trajectories are identified by setting appropriate initial conditions. Here, we explored initial conditions by setting $u = w = 0$ and incrementally increasing $q_v$ from $q_c$. Figure 6 illustrates an example of a 3D periodic trajectory and its projections onto the 2D phase plane.

To systematically trace a periodic trajectory and its bifurcation as a parameter (i.e., $\overline{q_v}$) changes, we used a mathematical technique known as numerical continuation. This method was implemented using the widely used software XPPAUT (Ermentrout and Mahajan 2003; Gandy and Nelson 2022). For a branch of solutions tracked from a periodic trajectory, Figure 7 presents changes in the wavelength (i.e., the distance between the moist clusters) and the minimum and maximum values of the three state variables with varying $\overline{q_v}$. The steady-state results obtained from numerical experiments on the time-dependent problem of the 1D globally coupled extended AN19 model are also presented.

To examine the characteristics of the solution branch, we start with $\overline{q_v} = 20\ \mathrm{kg\ m^{-2}}$ and first consider the wavelength (Fig. 7a). As $\overline{q_v}$ increases from $20\ \mathrm{kg\ m^{-2}}$, the wavelength decreases until $\overline{q_v} \approx 36\ \mathrm{kg\ m^{-2}}$. Beyond this point, the wavelength asymptotically increases to infinity as $\overline{q_v}$ approaches $q_{v0}$ (i.e., the uniform solution). Conversely, as $\overline{q_v}$ decreases from $20\ \mathrm{kg\ m^{-2}}$, the wavelength increases until $\overline{q_v} \approx 14\ \mathrm{kg\ m^{-2}}$. The branch folds back at $\overline{q_v} \approx 14\ \mathrm{kg\ m^{-2}}$, and the wavelength asymptotically increases to infinity as $\overline{q_v}$ approaches $\approx 16\ \mathrm{kg\ m^{-2}}$.

The results from the time-evolving numerical experiments generally agree with the solution branch obtained from the numerical continuation of the periodic trajectories of the dynamical system. Next, we consider the minimum and maximum values of $q_v$ (Fig. 7b). As $\overline{q_v}$ increases from $20\ \mathrm{kg\ m^{-2}}$, $q_v$-max decreases asymptotically to $q_{v0}$ as $q_v$ approaches $q_{v0}$. As $\overline{q_v}$ decreases from $20\ \mathrm{kg\ m^{-2}}$, $q_v$-max increases until $\overline{q_v} \approx 14\ \mathrm{kg\ m^{-2}}$. The branch folds back at $\overline{q_v} \approx 14\ \mathrm{kg\ m^{-2}}$, and $q_v$-max asymptotically increases to the upper stable fixed point of the reaction term as $\overline{q_v}$ approaches $\approx 16\ \mathrm{kg\ m^{-2}}$. As $\overline{q_v}$ increases from $20\ \mathrm{kg\ m^{-2}}$, $q_v$-min asymptotically increases to $q_v \approx 39\ \mathrm{kg\ m^{-2}}$ as $\overline{q_v}$ approaches $q_{v0}$. As $\overline{q_v}$ decreases from $20\ \mathrm{kg\ m^{-2}}$, after the branch folds back at $\overline{q_v} \approx 14\ \mathrm{kg\ m^{-2}}$, $q_v$-min asymptotically increases to the unstable fixed point of the reaction term as $\overline{q_v}$ approaches $\approx 16\ \mathrm{kg\ m^{-2}}$.

At $\overline{q_v} = q_{v0}$, the periodic trajectory asymptotically approaches a single fixed point, corresponding to a uniform solution. This phenomenon is known as a homoclinic bifurcation (or saddle-loop bifurcation) that is a type of infinite-period bifurcation (Strogatz 2019). Conversely, at $\overline{q_v} \approx 16\ \mathrm{kg\ m^{-2}}$, the periodic trajectory asymptotically approaches two fixed points in the phase plane of $q_v$ and $w$, as $w$ asymptotically approaches zero (Fig. 7d). These fixed points satisfy $\mathrm{d}q_v / \mathrm{d}x = 0$ and $\mathrm{d}w / \mathrm{d}x = 0$. The divergence of $u$ from infinity (Fig. 7c) supports this observation. This scenario represents a heteroclinic bifurcation. In summary, the wavelength can be infinitely long in a globally coupled system, and conversely, a nonlocally coupled system is required to allow for finite-scale selection of wavelengths.

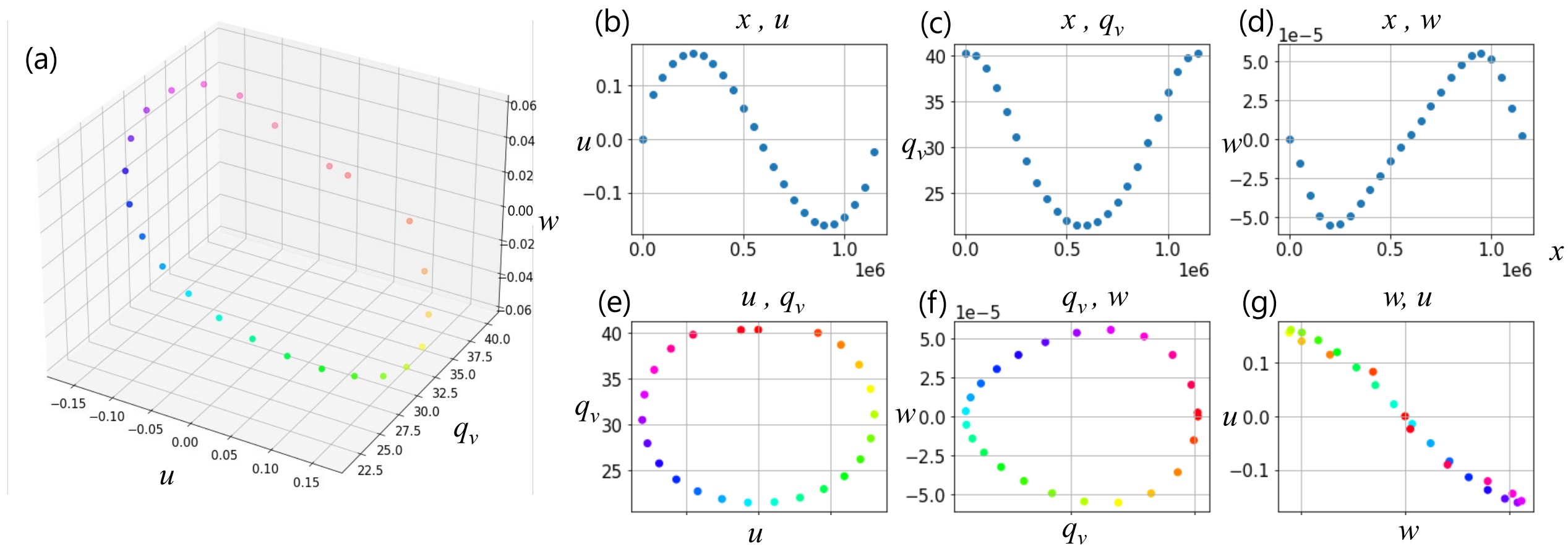

Fig. 6. An example of the trajectory and horizontal distribution of periodic solutions based on the dynamical system described by a set of Eq. (16). (a) Trajectory in a 3D phase space. (b)–(d) Horizontal variation of the state variables $u$, $q_v$, and $w$. (e)–(g) Trajectories of pairs of state variables projected onto a 2D phase plane.

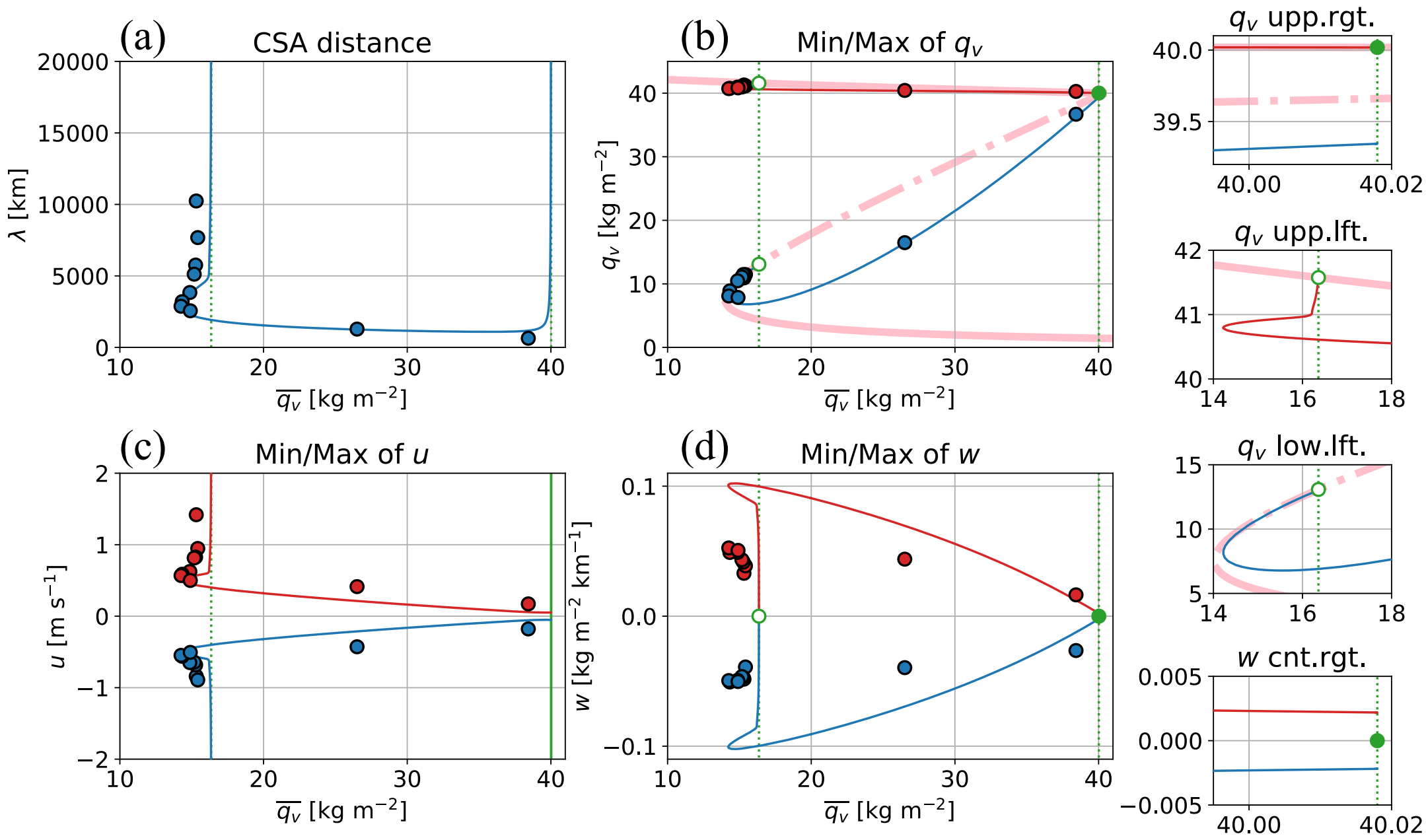

Fig. 7. Properties of periodic trajectories of the dynamical system describing the spatial evolution in the steady state of the globally coupled system as a function of $\overline{q_v}$. (a) The blue solid line represents the wavelength obtained through numerical continuation. Blue circles indicate the wavelength obtained from time-evolving numerical experiments. (b) Same as (a), except that red (blue) represents the maximum (minimum) of $q_v$ along the trajectories. (c) Same as (b), except for $u$. (d) Same as (b), except for $w$. Pink lines in (b) correspond to the black lines in Figure 2b. Green filled circles and the solid line represent the uniform solution ($q_v = q_{v0}$, $w = 0$, and $u$ can be any value). Green open circles represent the fixed points for the heteroclinic trajectory at $\overline{q_v} \approx 16$ kg m$^{-2}$; $q_v^*$ such that $\mathcal{R}(q_v^*) = 0$, $w = 0$. The rightmost small panels focus on sections of (b) and (d).

## 6. Discussion

Based on the present theory, how can CSA be linked to real tropical atmospheric phenomena? In this final section, we discuss the relationship between our theory, the moisture mode theory of the MJO, and water vapor variability in the tropical atmosphere. Sobel and Maloney (2012) analyzed east-west propagating and growing moisture modes along the equator and observed that the westward mode with the maximum growth rate appeared at a wavelength of 5,000 km, primarily due to surface flux feedback. They suggested that the instability arises from the asymmetry of the precipitation–wind response function, with the east-west wavelength depending on the characteristic scale of this response function. Furthermore, Sobel and Maloney (2013) reported the existence of an eastward-growing mode with a maximum growth rate at a wavelength of 40,000 km, noting that modes with scales of a few thousand kilometers were suppressed by additional factors such as synoptic disturbances that were not included in their earlier study.

The detailed mechanisms of our moisture model may differ from those in Sobel and Maloney's model. They attributed the wind speed dependence of surface fluxes and the east-west asymmetry of the precipitation–wind response function to destabilizing factors not included in our model. However, as Windmiller and Craig (2019) pointed out based on scaling arguments, there will likely be commonalities in spatial-temporal development even if specific feedback processes affecting moisture differ. Surface flux feedback can be integrated into our model, and we will explore its impact on the CSA mechanism in future studies. Notably, the precipitation-wind response function in their model was effectively a form of nonlocal coupling, and this aligns with our finding that nonlocal coupling is crucial for scale selection.

Beucler et al. (2019) compared the spatial variability of moisture across various settings: RCE experiments with CSA, a near-global experiment incorporating more realistic equatorial atmospheric dynamics, and reanalysis data corresponding to the real atmosphere. Their analysis suggested that vertical advection may have amplified the spectral peak at several thousand kilometers in CSA. This is consistent with the finding that the divergence term is a primary factor destabilizing the uniform state in our model. The absence of CSA-related peaks in the near-global experiment and real atmospheric spectra could be attributed to the effects of east-west asymmetry in waves and synoptic disturbances, as noted by Sobel and Maloney (2012, 2013).

In summary, we propose that CSA can be understood as an idealized form of tropical moisture organization, in which synoptic disturbances and zonal asymmetries associated with equatorial atmospheric dynamics are removed from the MJO framework based on moisture mode theory.

## 7. Conclusions

The present study proposes a nonlocally coupled moisture model that captures both the onset of CSA and horizontal scale selection. We first clarified analytically how bistable states are generated in the local dynamics of atmospheric columns in a moisture model based on the WTG approximation. Unlike previous studies that attempted to represent CSA by assuming an RD form of system equation with bistability, the model based on WTG framework naturally adds feedback from the circulation field excited by heating through an explicit advection term, in addition to reaction and diffusion terms. A previous study, with the goal of linking CSA to real-world phenomena, attempted to build a data-driven model of intertropical convergence zone assuming a priori RD form of system equation (Beucler et al. 2020). Based on our results, explicit inclusion of horizontal nonlocal coupling and advection effects would potentially improve model construction.

In this model, a linearly stable uniform state corresponds to a scattered convection regime. This uniform state can be destabilized by finite amplitude disturbances, transitioning to a non-uniform state that corresponds to an aggregated convection regime. For CSA to occur, the destabilizing effect of feedback from moisture–heating–circulation coupling must surpass the stabilizing effect of diffusion. This necessitates that the maximum wavelength, or the domain size for globally coupled systems, exceeds approximately 1,000 km. We analytically demonstrates that CSA onset is caused by the destabilization of a uniform state in response to a finite amplitude disturbance, and complements column-based framework by Emanuel et al. (2014) with horizontally extended field framework. This also provides a clear explanation for the hysteresis phenomenon of CSA onset reported in small-domain RCE simulations using CRMs (Muller and Held 2012; Jeevanjee and Romps 2013).

While the original globally coupled system exhibits its maximum effective growth rate at the longest wavelength, the newly proposed nonlocally coupled system exhibits a maximum effective growth rate at approximately the filter length scale. Although empirical filtering operations were used in this study, such nonlocal coupling plays essentially the same role as the heating response under the WTG approximation assumed in moisture mode theories (Sobel and Maloney 2012, 2013; Adames and Kim 2016).

In globally coupled systems, the wavelength can be infinitely extended, as the periodic trajectories in the dynamical system describing spatial evolution asymptotically approach a heteroclinic trajectory connecting local equilibrium solutions in moist and dry states. Additionally, the analytical method we used to elucidate this problem, termed numerical continuation, has not been used in atmospheric science before. The further application of such a mathematical method may help with a wide range of other theoretical studies in the future.

Finally, the limitations of this study and prospects are discussed. First, the present study could not account for the SST-dependence of CSA physics. While this study succeeded in incorporating the role of radiative interactions in the CSA onset into the mathematical model as mentioned in the introduction, the role of each physical process, including radiation, is not fully understood. For example, a recent numerical study reports that radiative interactions are not necessary for CSA onset in high-SST environment (Yao and Yang 2023), and this contrasts with a past theoretical study that has reported that high-SST is favorable for CSA owing to radiative interactions (Emanuel et al. 2014). Exploring a more general mathematical model that can comprehensively represent the SST dependence would be one of the next steps in advancing our theoretical understanding of CSA.

Second, the representation of the dynamic effect through the filter operation in the present model is empirical, too simple, and ignores the Earth's rotational effects. The finite filter length, which is a given parameter in our model, effectively determines the length of the CSA through the circulation field. Another framework to derive this length from a more principled law is an important subject for future work. While this study simply represents the GW adjustment process by a finite horizontal filtering in WTG balance equation, ideally, the response function should be constructed after explicitly treating the parameters of dynamical processes such as stratification effects and Earth rotation effects. In particular, if there is an Earth rotation effect, other length scales such as the Rossby deformation radius or the Rhines scale would need to be incorporated into the dynamical process (Chavas and Reed 2019). While our study discussed only simple situations in one horizontal dimension, it is necessary to consider organization in two horizontal dimensions (east-west and north-south) in the real tropical atmosphere. Some recent studies have investigated the intimate relationship among CSA, moisture mode, and large-scale tropical circulation fields on the $f$-plane or equatorial $\beta$-plane (Davison and Haynes 2024; Adames-Corraliza and Mayta 2024). While cloud self-organization studies have been developed from small to large scales using nonhydrostatic equation models (e.g., Nakajima and Matsuno 1988; Bretherton et al. 2005; Yanase et al. 2022b), large-scale tropical atmospheric dynamics studies have been developed by incorporating moist physics into shallow water equations or primitive equation models (e.g., Matsuno 1966; Gill 1980; Neelin and Yu 1994; Sobel et al. 2001; Adames and Kim 2016). The two branches are now asymptotically approaching unification. Our work extends the scope of conventional moisture mode theory framework, making it applicable not only to planetary-scale phenomena such as the MJO but also to mesoscale organized tropical convection phenomena, including CSA. Additionally, the development of a simple mathematical model framework provides us with not only a theoretical understanding but also a perspective to grasp the factors that contribute to the discrepancy in results between different complex models.

*Acknowledgments.*

The authors thank Yoshiyuki Kajikawa, Yuta Kawai, and Hiroaki Miura for insightful discussions. This study was partly supported by JSPS KAKENHI Grant Number JP24K17128. T.Y. received support from the RIKEN Special Postdoctoral Researcher Program.

*Data Availability Statement.*

The source codes for the simulation and analysis are locally stored at the RIKEN R-CCS.

APPENDIX

**Details of the nonlocally coupled moisture model**

*a. Basic equations*

We have largely followed Ahmed and Neelin (2019). One of the main differences from the original system of equations is the exclusion of water condensate as a prognostic variable and the associated cloud radiative effect. This facilitates a semi-analytical understanding. Additionally, the rotational component of horizontal wind is excluded. This change is made because we are interested in the behavior of clouds in nonrotating systems, as is typically adopted in the framework of RCE. Other changes are made to the vertical distribution functions and other parameters. More reproducible analytically simple expressions are used for the vertical distribution functions rather than those aligned with real atmospheric observations. Although these parameters may alter the quantitative results, they do not affect the main conclusions. For example, changes in the phase diagram across a wide range of parameters can be easily confirmed. In the following section, the specific system of equations is explained. The variables, functions, and parameters used in this study are summarized in Tables A1 and A2.

*b. Mass continuity equation*

The mass continuity equation in pressure coordinates is:

$$\frac{\partial \omega}{\partial p} + \nabla \cdot \boldsymbol{V} = 0, \tag{A1}$$

where $\omega = \omega(x, y, p, t)$ [Pa s$^{-1}$] is the vertical velocity, $p$ [Pa] is the pressure, and $\boldsymbol{V} = \boldsymbol{V}(x, y, p, t)$ [m s$^{-1}$] is the horizontal velocity vector.

We assume that the vertical velocity has a prescribed vertical structure:

$$\omega = -a\Omega_1, \tag{A2}$$

where $a = a(x, y, t)$ [$\mathrm{s^{-1}}$] is the horizontally and temporally varying component, the horizontal divergence in the upper troposphere, and $\Omega_1 = \Omega_1(p)$ [Pa] is the vertically varying component, the vertical structure function.

According to Equations (A1) and (A2):

$$\nabla \cdot \boldsymbol{V} = \frac{\mathrm{d}\Omega_1}{\mathrm{d}p} \nabla \cdot \boldsymbol{v},$$

where $\boldsymbol{v} = \boldsymbol{v}(x, y, t)$ [$\mathrm{m\ s^{-1}}$] is the horizontal divergent wind such that $\nabla \cdot \boldsymbol{v} = a$. In this study, the vertical structure of vertical velocity is expressed as follows:

$$\Omega_1 = \frac{p_\mathrm{b} - p_\mathrm{t}}{2} \sin\left(\frac{\pi(p - p_\mathrm{t})}{p_\mathrm{b} - p_\mathrm{t}}\right),$$

where $p_\mathrm{b}$[Pa] and $p_\mathrm{t}$ [Pa] are the pressures at the bottom and top of the atmospheric column (constant), respectively. This is generally referred to as the first baroclinic mode.

*c. Thermodynamic equation*

We consider a vertically integrated dry static energy budget equation. We applied the WTG approximation, indicating that the local tendency term and horizontal advection terms were neglected (Sobel et al. 2001; Sobel and Bretherton 2000; Sobel et al. 2007; Ruppert and Hohenegger 2018; Yanase et al. 2022a). Convective and radiative heating were considered for horizontally and temporally varying vertically integrated diabatic heating. The sensible heat flux at the ground surface was assumed to be constant. Additionally, a stochastic term is included.

The diagnostic balance equation of dry static energy $s$ [$\mathrm{J\ kg^{-1}}$] is modeled as:

$$\left\langle \omega \frac{\mathrm{d}s}{\mathrm{d}p} \right\rangle = L_v\left(\mathcal{P} - \tilde{\mathcal{P}}\right) + \left(\mathcal{F}_\mathrm{rad} - \tilde{\mathcal{F}}_\mathrm{rad}\right) - \xi,$$

where the left-hand side represents the vertically integrated vertical advection of dry static energy, $L_v$ [$\mathrm{W\ m^{-2}\ kg^{-1}\ m^2\ s}$] is the latent heat constant, $\mathcal{P} = \mathcal{P}(q_v)$ [$\mathrm{kg\ m^{-2}\ s^{-1}}$] is the surface precipitation rate as a function of vertically integrated water vapor content $q_v = q_v(x, y, t)$, $\mathcal{F}_\mathrm{rad} = \mathcal{F}_\mathrm{rad}(q_v)$ [$\mathrm{W\ m^{-2}}$] is the column radiative heating rate as a function of $q_v$, and $\xi = \xi(x, y, t)$ [$\mathrm{W\ m^{-2}}$] is the stochastic term described by the Ornstein–Uhlenbeck process.

The expression $\langle A \rangle$ denotes the vertical integral of variable $A$ as follows:

$$\langle A \rangle = \frac{1}{g} \int_{p_\mathrm{t}}^{p_\mathrm{b}} A(p') \mathrm{d} p',$$

where $g$ [m s$^{-2}$] is the gravitational acceleration constant. The expression $\tilde{A}$ represents the filtered value of variable $A$ (see the main manuscript). We assume that the change in dry static energy with respect to pressure is constant (i.e., $\mathrm{d}s/\mathrm{d}p = \mathrm{const.}$).

By using this specified vertical structure function and integrating the vertical advection term:

$$\nabla \cdot \boldsymbol{v} = \frac{L_v\left(\mathcal{P} - \tilde{\mathcal{P}}\right) + \left(\mathcal{F}_\mathrm{rad} - \tilde{\mathcal{F}}_\mathrm{rad}\right) - \xi}{M_s}, \qquad \text{(A3)}$$

where $M_s = (\mathrm{d}s/\mathrm{d}p)(p_\mathrm{b} - p_\mathrm{t})^2/(g\pi)$ is referred to as the gross dry stability.

The surface precipitation rate $\mathcal{P}$ is parameterized as a piecewise linear function of $q_v$:

$$\mathcal{P} = \alpha(q_v - q_c)\mathcal{H}(q_v - q_c)$$

where $\alpha$ is the coefficient of relaxation by precipitation, $q_c$ is the critical column water vapor content, and $\mathcal{H}$ is the Heaviside step function such that $\mathcal{H}(A) = 0$ for $A \leq 0$ and $\mathcal{H}(A) = 1$ for $A > 0$.

The radiative heating rate $\mathcal{F}_\mathrm{rad}$ is parameterized as a linear function of $q_v$:

$$\mathcal{F}_\mathrm{rad} = \epsilon_\mathrm{r} q_v + \mathrm{const.}$$

where $\epsilon_\mathrm{r}$ [W m$^{-2}$ kg$^{-1}$ m$^2$] is the coefficient of the water vapor radiative effect.

The stochastic term $\xi$, reflecting the heating variability not accounted for by precipitation and radiation, is parameterized by the following stochastic time evolution equation:

$$\mathrm{d}\xi = -\tau_\mathrm{noise}^{-1} \xi \mathrm{d}t + \tau_\mathrm{noise}^{-1} \sigma_\mathrm{noise} \mathrm{d}W(t)$$

where $\tau_\mathrm{noise}$ is the damping time scale, $\sigma_\mathrm{noise}$ is the intensity of variability, and $\mathrm{d}W(t)$ is the Wiener process.

*d. Water vapor equation*

We assume that the specific humidity $q$ [kg kg$^{-1}$] possesses a prescribed vertical structure:

$$q = q_\mathrm{v} \Omega_2,$$

where $q_v = \langle q \rangle$ is the vertically integrated water vapor content, and $\Omega_2 = \Omega_2(p)$ [$\mathrm{Pa^{-1}\ m\ s^{-2}}$] expresses the vertical variation of specific humidity:

$$\Omega_2(p) = \frac{g\left(e^{p/p_q} - e^{p_\mathrm{t}/p_q}\right)}{p_\mathrm{q}\left(e^{p_\mathrm{b}/p_q} - e^{p_\mathrm{t}/p_q}\right) - e^{p_\mathrm{t}/p_q}(p_\mathrm{b} - p_\mathrm{t})}$$

where $p_q$ [Pa] is the scale of the pressure for a vertical change in water vapor content. This function form is chosen such that it satisfies $\Omega_2(p_\mathrm{t}) = 0$, $\langle \Omega_2 \rangle = 1$, and increases exponentially with pressure.

The prognostic equation of vertically integrated water vapor content $q_v$ is modeled as:

$$\frac{\partial q_v}{\partial t} = E - \mathcal{P} + M_q(\nabla \cdot \boldsymbol{v})q_v + M_q(\boldsymbol{v} \cdot \nabla)q_v + D\nabla^2 q_v,$$

where $E$ [$\mathrm{kg\ m^{-2}\ s^{-1}}$] is the evaporation rate constant, $D$ [$\mathrm{m^2\ s^{-1}}$] is the eddy diffusion coefficient constant, and $M_q = \langle -\Omega_2 \mathrm{d}\Omega_1/\mathrm{d}p \rangle$ is referred to as the gross moisture stratification.

Additionally, by introducing the velocity potential $\phi$ such that $\nabla^2\phi$ equals the rhs of Equation (A3) and solving Poisson's equation, the spatial field of $\boldsymbol{v}$ is determined.

The reaction term defined in the main manuscript is a quadratic function of $q_v$ below and above $q_c$, respectively. For $q_\mathrm{v} \leq q_\mathrm{c}$:

$$\mathcal{R} = A_1 q_v^2 + B_1 q_v + C_1, \tag{A4}$$

$$\text{where } A_1 = \frac{M_q}{M_s}\epsilon_r,\ B_1 = -\frac{M_q}{M_s}\left(L_v\tilde{\mathcal{P}} + \epsilon_r\widetilde{q_v}\right),\ C_1 = E.$$

For $q_\mathrm{v} > q_c$:

$$\mathcal{R} = A_2 q_v^2 + B_2 q_v + C_2, \tag{A5}$$

$$\text{where } A_2 = \frac{M_q}{M_s}(\alpha L_v + \epsilon_r),\ B_2 = -\left(\alpha + \alpha\frac{M_q}{M_s}L_v q_c + \tilde{\mathcal{P}}\frac{M_q}{M_s}L_v + \frac{M_q}{M_s}\epsilon_r\widetilde{q_v}\right),\ C_2 = E + \alpha q_c.$$

The potential function is defined as:

$$V(q_v) = -\int_0^{q_v} R(q_v{}')\mathrm{d}\,q_v{}' + \text{const.}, \tag{A6}$$

where the constant is set such that $\mathcal{V}(q_c) = 0$. Fig. A1 (a) and (b) provides examples of the tendency and potential as functions of the column water vapor content for a given radiative coefficient and different average column water vapor content values. Figs. A1 (c)–(f) also indicates the discriminants for the quadratic equations of Equations (A4) and (A5) being zero, respectively, and the solutions $Q_1, Q_2, Q_3,$ and $Q_4$, that are the local equilibrium points, as functions of the average column water vapor content and the radiative coefficient.

*e. Conditions for multiple equilibria*

In this section, we consider a dynamic system described solely by the reaction term, as discussed in Fig. 2, and we present the conditions for a multiple equilibrium state. The reaction term is a quadratic function that varies depending on whether $q_v$ is greater than or less than $q_c$. Therefore, we analyze the system in different cases. The quadratic function is convex downward in each range (i.e., $A_1 > 0$ and $A_2 > 0$), and the maximum number of stable solutions is two. Consequently, multiple equilibrium states are bistable.

The condition for the existence of a dry stable solution is that the quadratic equation given in Eq. (A4) must have two real solutions, $Q_1$ and $Q_2 (> Q_1)$, and the stable solution $Q_2$ must be within the domain. Specifically, this requires $B_1^2 - 4A_1C_1 > 0$ and $0 < Q_2 < q_c$. Similarly, the condition for the existence of a moist stable solution is that the quadratic equation given in Eq. (A5) must have two real solutions, $Q_3$ and $Q_4$ $(Q_4 > Q_3)$. This requires $B_2^2 - 4A_2C_2 > 0$ and $Q_3 > q_c$.

The parameter region where two stable solutions exist is indicated in Fig. A1 that presents the dependence of $B_1^2 - 4A_1C_1, Q_1$, $Q_2$, $B_2^2 - 4A_2C_2$, $Q_3$, and $Q_4$ on $\epsilon_{\mathrm{r}}$ and $\widetilde{q_v}$.

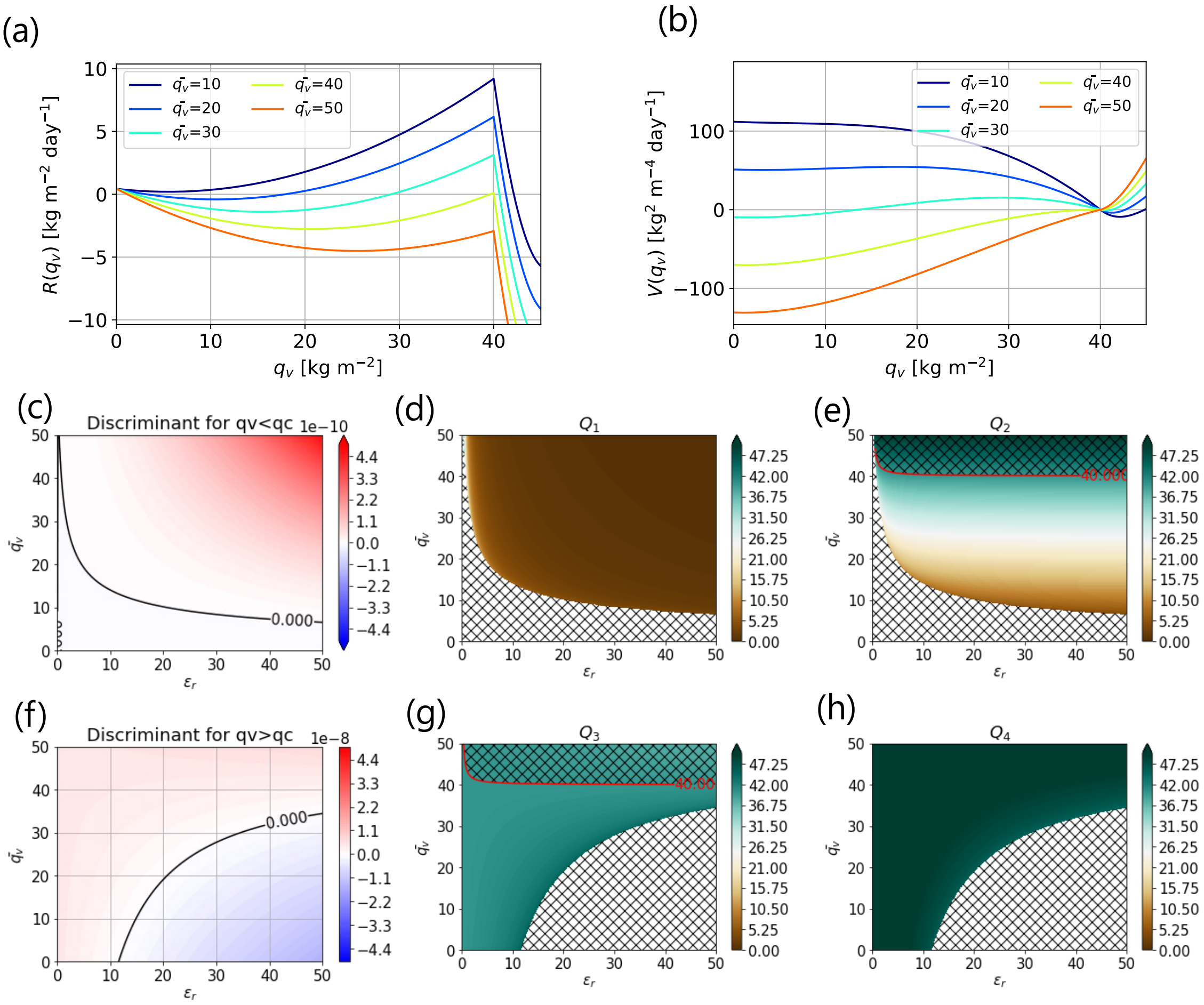

Fig. A1. Characteristics of the reaction term. (a) Tendency of the reaction term and (b) potential function as a function of column water vapor content $q_v$ for various average column water vapor contents $\widetilde{q_v}$. The case presented is for $\epsilon_r = 10$ W m$^{-2}$ kg$^{-1}$ m$^2$. (c) Discriminant, (d)–(e) solutions for the quadratic equation of the reaction term in Eq. (A4) set to zero, and (f) discriminant, (g)–(h) solutions for the quadratic equation of the reaction term in Eq. (A5) set to zero, as functions of $\epsilon_r$ and $\widetilde{q_v}$.

| Symbol | Description | Unit | Note |
|---|---|---|---|
| $p$ | Pressure | Pa | - |
| $\omega$ | Vertical wind velocity | $\mathrm{Pa\ s^{-1}}$ | - |
| $\boldsymbol{V}$ | Horizontal wind velocity vector | $\mathrm{m\ s^{-1}}$ | - |
| $\boldsymbol{v}$ | Horizontal divergent velocity vector in the upper troposphere | $\mathrm{m\ s^{-1}}$ | - |
| $s$ | Dry static energy | $\mathrm{J\ kg^{-1}}$ | - |
| $q$ | Specific humidity | $\mathrm{kg\ kg^{-1}}$ | - |
| $q_v$ | Column water vapor content | $\mathrm{kg\ m^{-2}}$ | - |
| $\xi$ | Stochastic heating | $\mathrm{W\ m^{-2}}$ | - |
| $\mathcal{P}$ | Surface precipitation rate | $\mathrm{kg\ m^{-2}\ s^{-1}}$ | Function of $q_v$ |
| $\mathcal{F}_{\mathrm{rad}}$ | Column radiative heating rate | $\mathrm{W\ m^{-2}}$ | Function of $q_v$ |
| $\mathcal{H}$ | Heaviside step function | - | - |

Table A1. List of variables and functions

| Symbol | Description | Value | Note |
| --- | --- | --- | --- |
| $E$ | Evaporation rate | $5.0 \times 10^{-6}$ kg m$^{-2}$ s$^{-1}$ | - |
| $D$ | Eddy diffusion coefficient | $7.5 \times 10^{4}$ m$^{2}$ s$^{-1}$ | - |
| $M_q$ | Gross moisture stratification per unit $q_v$ | 1.14 | - |
| $p_q$ | Pressure scale of water vapor change | 20,000 Pa | - |
| $M_s$ | Gross dry stability | $1.3 \times 10^{8}$ J m$^{-2}$ | $\mathrm{d}s/\mathrm{d}p$ = –0.5 J kg$^{-1}$ Pa$^{-1}$ |
| $L_v$ | Latent heat | $2.16 \times 10^{6}$ W m$^{-2}$ kg$^{-1}$ m$^{2}$ s | Column heating per precipitation rate |
| $g$ | Gravitational acceleration | 9.81 m s$^{-2}$ | - |
| $p_\mathrm{b}$ | Pressure at the troposphere bottom | 100,000 Pa | - |
| $p_\mathrm{t}$ | Pressure at the troposphere top | 10,000 Pa | - |
| $\alpha$ | Coefficient of water vapor relaxation by precipitation | 3,600$^{-1}$s$^{-1}$ | - |
| $q_c$ | Critical column water vapor content | 40 kg m$^{-2}$ | - |
| $\epsilon_\mathrm{r}$ | Coefficient of water vapor radiative effect | 10 W m$^{-2}$ kg$^{-1}$ m$^{2}$ | - |
| $\tau_\mathrm{noise}^{-1}$ | Damping time scale of the stochastic process | 7,200 s | - |
| $\sigma_\mathrm{noise}$ | Amplitude of stochastic forcing | 30 W m$^{-2}$ | - |

Table A2. List of parameters